\documentclass[PAPER, atlasdraft=false, USenglish, cernpreprint, texmf, orcidlogo]{atlasdoc}

\usepackage{atlaspackage}
\usepackage{atlasbiblatex}
\usepackage[process,misc]{atlasphysics}

\graphicspath{{logos/}{figures/}}

\providecommand{\mgg}{\ensuremath{m_{\gamma\gamma}}\xspace}
\providecommand{\pTgg}{\ensuremath{p_{\mathrm{T}}^{\gamma\gamma}}\xspace}
\providecommand{\GH}{\ensuremath{\Gamma_H}\xspace}
\providecommand{\GHSM}{\ensuremath{\Gamma_H^{\mathrm{SM}}}\xspace}

\providecommand{\Sherpa}{\SHERPA}

\AtlasTitle{Constraints on the Higgs boson total width from on-shell signal--background interference in the $H\to\gamma\gamma$ decay channel with $pp$ collisions at $\sqrt{s}$~=~13~TeV with the ATLAS detector}

\AtlasAbstract{%
The total decay width of the Higgs boson, $\Gamma_H$, encodes all of its decay modes; any decay into particles beyond the Standard Model (SM) would increase $\Gamma_H$ above the SM prediction of 4.1 MeV. In this Letter, a direct, on-shell measurement of $\Gamma_H$ is reported, exploiting the interference between the resonant $H \to \gamma\gamma$ amplitude and the continuum diphoton background, which distorts the diphoton invariant mass distribution in a way that depends on $\Gamma_H$ without requiring assumptions on the Higgs boson couplings. Using 140~$\text{fb}^{-1}$ of proton--proton collision data at $\sqrt{s}$~=~13~TeV recorded by the ATLAS detector during Run~2 of the LHC, an observed (expected) upper limit of $\Gamma_H$~<~307~(258)MeV is set at 95\% confidence level, approximately 75~(63) times the SM prediction. Unlike indirect off-shell constraints on $\Gamma_H$, this measurement does not rely on assumptions about the Higgs boson couplings, providing a novel complementary probe sensitive to potential Higgs boson decays into new particles.
}

\AtlasRefCode{ANA-HIGP-2024-12}

\PreprintIdNumber{CERN-EP-2026-261}

\AtlasJournal{Phys.\ Rev.\ Lett.}

\hypersetup{pdftitle={ATLAS paper ANA-HIGP-2024-12},pdfauthor={The ATLAS Collaboration}}

\begin{document}

\maketitle

After the discovery of the Higgs boson by the ATLAS and CMS collaborations at the Large Hadron Collider (LHC) in 2012~\cite{HIGG-2012-27,CMS-HIG-12-028}, the precise determination of its properties became one of the primary goals of their physics programmes~\cite{HIGG-2021-23,CMS-HIG-22-001}. The total Higgs boson decay width \(\Gamma_H\) is of particular interest: it is the sum of all partial decay widths, and any decay mode into new, undetected particles would increase \(\Gamma_H\) above the Standard Model (SM) prediction of \(\Gamma_H^{\text{SM}} = 4.1\)~\MeV at Higgs boson mass \(m_H = 125\)~\GeV~\cite{deFlorian:2016spz}. A determination of \(\Gamma_H\) therefore provides a test of the SM that is sensitive to a broad range of extensions of the Higgs sector~\cite{Curtin:2013fra}, including decays into new particles that could be invisible or otherwise undetected~\cite{HIGG-2021-05,CMS-HIG-21-007}.

The predicted natural width of the Higgs boson is about two to three orders of magnitude smaller than the experimental mass resolution. This resolution is $\mathcal{O}(1\text{--}3)$~\GeV for both the diphoton ($m_{\gamma\gamma}$) invariant mass~\cite{HIGG-2019-16} and the four-lepton invariant mass in $H \to ZZ^{*} \to 4\ell$~\cite{HIGG-2020-07}, the two fully reconstructed final states with the best mass resolution. As a result, the width of the Higgs boson cannot be directly resolved from the reconstructed lineshape. Indirect constraints on \(\Gamma_H\) are obtained by comparing the Higgs boson production rates near the mass peak (on-shell) and far above it (off-shell) in the \(H \to ZZ\)~\cite{HIGG-2018-32,HIGP-2024-14,CMS-HIG-21-013,CMS-HIG-21-019} and \(H \to WW\)~\cite{HIGP-2024-05,CMS-HIG-24-011} decay channels. The most precise determinations are obtained from their combination: the ATLAS experiment measures \(\Gamma_H = 3.6^{+2.1}_{-1.6}\)~\MeV, combining the off-shell measurements with the on-shell Higgs boson production measurements~\cite{HIGP-2025-19}, while the CMS experiment obtains \(\Gamma_H = 5.1^{+2.0}_{-1.8}\)~\MeV~\cite{CMS-HIG-25-003}. The interpretation of these results relies on the assumption that the Higgs boson couplings to gluons and vector bosons are identical in on-shell and off-shell production, as discussed in Ref.~\cite{Englert:2014aca}.

A complementary and novel approach, used in this Letter, exploits the interference between the resonant Higgs boson amplitude and the non-resonant continuum amplitude~\cite{Dixon:2003,Martin:2012,Dixon:2013,deFlorian:2013,Martin:2013jet,Coradeschi:2015,Campbell:2017,Bargiela:2023}: since both processes share the same initial and final states, the two amplitudes add coherently, giving rise to an interference term when the total amplitude is squared. The dominant effect arises from interference between the gluon-induced $gg\rightarrow H\rightarrow \gamma \gamma$ resonance and the continuum $gg\rightarrow \gamma\gamma$ background. A contribution roughly an order of magnitude smaller and of opposite sign arises from quark--gluon scattering, in which the \(qg \to Hq \to \gamma\gamma q\) amplitude interferes with the \(qg \to \gamma\gamma q\) continuum~\cite{Martin:2013jet,Coradeschi:2015}. The interference term has two components. The first, proportional to the real part of the product of the signal and background amplitudes, is an antisymmetric function of $m_{\gamma \gamma}^2$ around the Higgs boson resonance. It produces an imbalance of events on either side of the mass peak that is equivalent to an apparent shift of the peak position of approximately $-70$~\MeV at the SM width~\cite{Martin:2012,Bargiela:2023,Buccioni:2026}.  The second component, proportional to the imaginary part, is symmetric around the peak but modifies the overall signal yield by approximately $-1.7\%$ at the SM width~\cite{Bargiela:2023}. The dependence of these effects on \(\Gamma_H\) can be made explicit following Refs.~\cite{Dixon:2013,Bargiela:2023}.  Expressing the $gg$-induced signal rate relative to its SM prediction through the signal-strength parameter \(\mu\), the interference scales as \(I/I_{\text{SM}} = \sqrt{\mu\,\Gamma_H/\Gamma_H^{\text{SM}}}\), where $I_{\text{SM}}$ is the interference contribution at the SM width.  By profiling $\mu$, the measurement is independent of the values of the Higgs boson couplings and of the mechanism responsible for a possible enhancement of the total width, such as decays into undetected particles.

\newcommand{\AtlasCoordFootnote}{%
ATLAS uses a right-handed coordinate system with its origin at the nominal interaction point (IP) in the centre of the detector and the \(z\)-axis along the beam pipe.  The \(x\)-axis points from the IP to the centre of the LHC ring, and the \(y\)-axis points upwards.  Polar coordinates \((r,\phi)\) are used in the transverse plane, \(\phi\) being the azimuthal angle around the \(z\)-axis.  The pseudorapidity is defined in terms of the polar angle \(\theta\) as \(\eta = -\ln \tan(\theta/2)\) and is equal to the rapidity $ y = \frac{1}{2} \ln \left( \frac{E + p_z}{E - p_z} \right) $ in the relativistic limit.  Angular distance is measured in units of \(\Delta R \equiv \sqrt{(\Delta y)^{2} + (\Delta\phi)^{2}}\).}

This Letter reports a measurement of \(\Gamma_H\) in the diphoton channel using the full \(pp\) collision data collected at \(\sqrt{s}=13~\TeV\) by the ATLAS detector during Run~2 of the LHC (2015--2018). The sensitivity to \GH comes mainly from the interference-induced change of the \(m_{\gamma\gamma}\) lineshape, which shifts the peak towards lower \mgg values, and to a lesser extent from the reduction of the signal yield relative to the SM prediction. Both effects scale as $\sqrt{\GH}$. An external constraint on the Higgs boson mass from the $H\to ZZ^*\to 4\ell$ measurement~\cite{HIGG-2020-07} is a key input to the analysis, since the interference-induced mass shift is measured relative to the measured value of $m_H$. A similar measurement has recently been reported by the CMS Collaboration~\cite{CMS-HIG-25-004}; the difference in sensitivity relative to the analysis presented in this Letter mainly reflects the more conservative treatment adopted here for the theory uncertainties and the interference modelling.

The ATLAS experiment~\cite{PERF-2007-01} is a multipurpose particle detector with a forward--backward symmetric cylindrical geometry and a near \(4\pi\) coverage in solid angle.\footnote{\AtlasCoordFootnote} It consists of an inner tracking detector surrounded by a thin superconducting solenoid, electromagnetic and hadronic calorimeters, and a muon spectrometer.  A two-level trigger system was used to select events~\cite{TRIG-2016-01}; the events used in this analysis were recorded using diphoton triggers with transverse-energy thresholds of 35 and 25~\GeV~\cite{TRIG-2018-05} for the leading and subleading photon candidates, respectively.  The integrated luminosity of the data sample after trigger and data-quality requirements is $140.1 \pm 1.2~\ifb$~\cite{DAPR-2021-01}.

The $H\to\gamma\gamma$ signal process and the $gg$- and $qg$-initiated interference contributions were generated with \Sherpa~2.2.11~\cite{Bothmann:2019yzt}, using a dedicated \Sherpa\ module that simulates both interference processes.  The matrix elements are accurate to next-to-leading-order (NLO) in quantum chromodynamics (QCD) for the signal and the $gg$-interference, and to leading order (LO) for the $qg$-interference~\cite{Bern:2002jx, Dixon:2013}, and are matched to the Catani--Seymour dipole parton shower (CSS)~\cite{Schumann:2007mg} implemented in \Sherpa.  The Higgs boson mass was set to $m_H = 125~\GeV$, and samples were produced at $\Gamma_H = \Gamma_H^{\text{SM}}$ and $\Gamma_H = 100 \times \Gamma_H^{\text{SM}}$. The latter is used exclusively in the injection tests described below.  All signal and interference samples were passed through the full \GEANT-based simulation of the ATLAS detector~\cite{Agostinelli:2002hh,SOFT-2010-01}.  The signal yield is normalised with a correction factor $K_S = 1.4$, accounting for higher-order QCD corrections up to next-to-next-to-next-to-leading-order (N$^3$LO)~\cite{Anastasiou:2016cez}, and the $gg$-interference yield with $K_I = 1.64$, corresponding to next-to-next-to-leading-order (NNLO) in the soft-virtual approximation (NNLOsv)~\cite{Bargiela:2023,Buccioni:2026}.  The corresponding NLO correction to the $qg$-interference yield is found to be negligible and no additional $K$-factor is applied.  All samples were generated using the NNPDF3.0 NNLO parton distribution functions~\cite{Ball:2014uwa}.  The other Higgs boson production modes are included as non-interfering signal contributions.  They are simulated with the \POWHEGBOX~\cite{Alioli:2010xd} event generators interfaced to \PYTHIA[8]~\cite{Sjostrand:2014zea}, and are normalised to state-of-the-art SM calculations of the production cross-sections and branching ratios~\cite{deFlorian:2016spz}.  Background events from non-resonant diphoton production were simulated with \Sherpa~2.2.11, with matrix elements at NLO accuracy for up to one additional parton and at LO for two or three additional partons, and processed with a fast parameterised simulation of the calorimeter response~\cite{SOFT-2010-01}; since the non-resonant background is determined from data, the simulation is used only to select and validate the analytic background model.

The event selection closely follows that of the ATLAS $H\to\gamma\gamma$ analyses based on the full Run~2 data sample~\cite{HIGG-2020-16,HIGG-2019-16}.  Events are required to contain at least two photon candidates with pseudorapidity $|\eta| < 2.37$, excluding the barrel--endcap calorimeter transition region $1.37 < |\eta| < 1.52$, meeting \emph{tight} identification and isolation criteria~\cite{EGAM-2021-01} and matched to the online photon candidates that satisfied the trigger selection.  The diphoton production vertex is selected with a neural-network algorithm~\cite{HIGG-2013-08}.  The transverse energies, $E_\mathrm{T}$, of the $E_\mathrm{T}$-leading and $E_\mathrm{T}$-subleading photon candidates are required to exceed $0.35\,\mgg$ and $0.25\,\mgg$, respectively, and the diphoton invariant mass must satisfy $105 < \mgg < 160~\GeV$.

The selected events are classified into two mutually exclusive categories according to the diphoton transverse momentum, $\pTgg$.  The low-$\pTgg$ category contains events with $\pTgg < 30~\GeV$, where the size of the interference contribution relative to the signal is largest; the high-$\pTgg$ category contains events with $\pTgg \geq 30~\GeV$, with slightly smaller interference effects.  A finer categorisation is not adopted because, although the signal and $gg$-interference matrix elements are NLO-accurate in QCD, the $\pTgg$ spectrum itself is only LO-accurate, since a non-zero $\pTgg$ arises only through the real-emission correction. Moreover, at higher $\pTgg$ the predictions of the interference obtained with different parton-shower algorithms diverge significantly~\cite{Bothmann:2019yzt,Schumann:2007mg}. Finally, the bulk of the $gg$-interference contribution populates the low-$\pTgg$ region, so the number of available Monte Carlo (MC) simulated events diminishes rapidly towards higher $\pTgg$. All these effects make the modelling insufficiently precise to support a finer $\pTgg$ categorisation, motivating the use of only two regions separated by a single $\pTgg$ boundary.  The boundary is optimised to minimise the expected upper limit on \GH, and lies close to the value proposed in Ref.~\cite{Dixon:2013}.  An improved categorisation will be possible once higher-order predictions for the interference matched to parton showers become available.

Figure \ref{fig:mass_shift} compares the signal-only $m_{\gamma\gamma}$ distributions with the combined signal-plus-interference distributions per category, such that the resulting shift in the mass peak is evident. The lower panel of Figure~\ref{fig:mass_shift} shows the absolute difference between the signal-plus-interference and signal-only distributions, in each category; the characteristic antisymmetric shape, positive below $m_H$ and negative above it, reflects the mass-peak shift towards lower $m_{\gamma\gamma}$ values induced by the interference. The larger magnitude of the difference in the low-$\pTgg$ category is consistent with the dominant $gg$-interference contribution being enhanced at low $\pTgg$.

\begin{figure}[tp]
\centering
\includegraphics[width=0.49\linewidth]{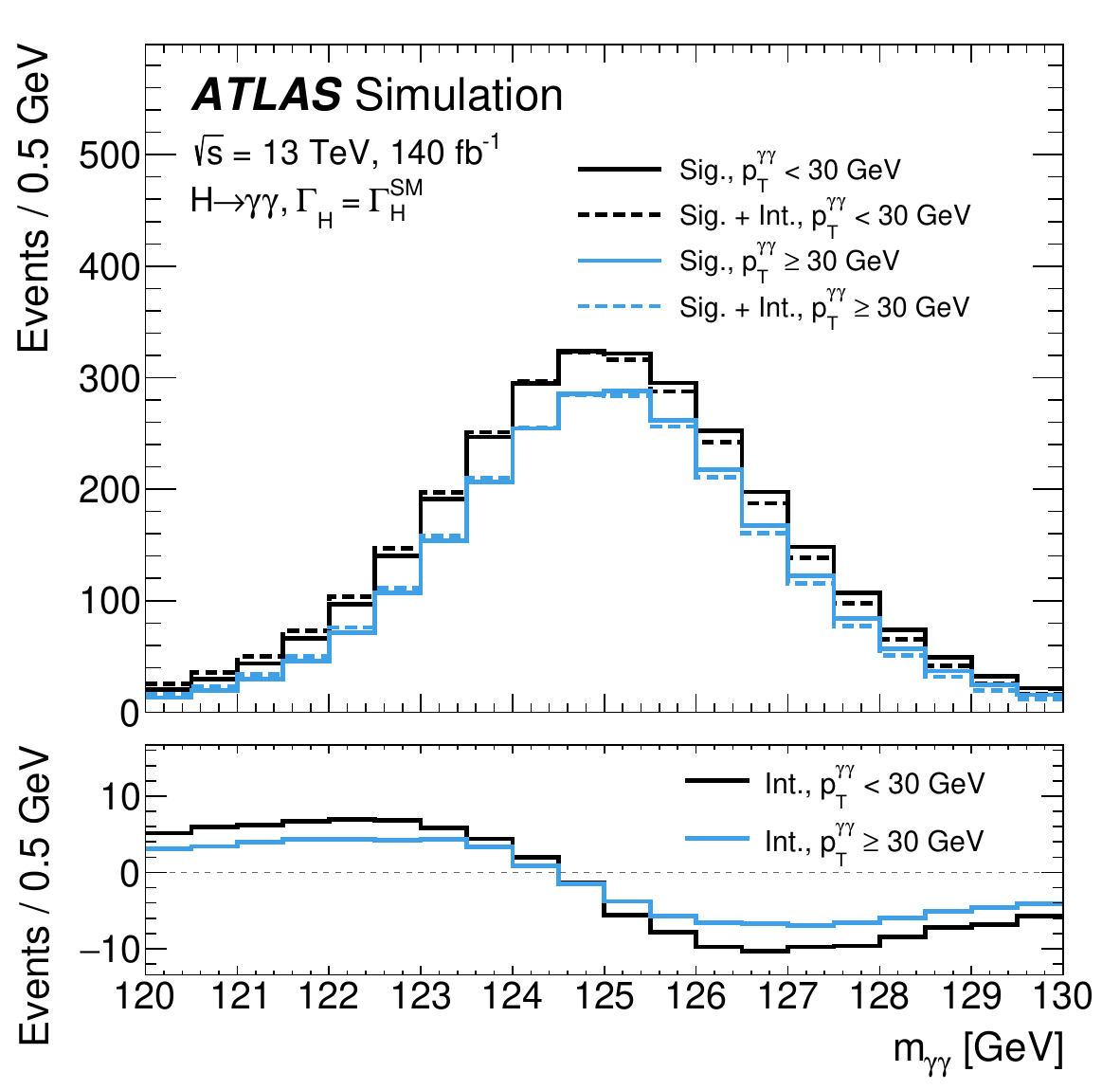}
\captionsetup{width=0.9\textwidth}
\caption{Distributions of $m_{\gamma\gamma}$ for the signal-only (solid) and
signal-plus-interference (dotted) predictions in the $\pTgg < 30~\GeV$ (black) and
$\pTgg \geq 30$~\GeV (light blue) categories, shown in the region $120 < m_{\gamma\gamma} < 130$~\GeV
at the SM width. The lower panel shows the interference-only distributions, illustrating the induced distortion of the
signal lineshape in each category.} \label{fig:mass_shift}
\end{figure}

The Higgs boson total width is determined from a simultaneous maximum-likelihood fit to the \mgg distributions in the two \pTgg categories, with analytic models for the signal, interference and background described below.

The \mgg distribution of the \Hgg signal in each category is found to be well described by a Double-Sided Crystal Ball (DSCB) function~\cite{Oreglia:1980cs,HIGG-2014-04}, consisting of a Gaussian core with power-law tails on both sides, whose six free parameters are determined from a fit to the simulated \mgg distribution at the SM width.  The dependence of the shape parameters on \GH is negligible within the sensitivity of the analysis, and the parameters are therefore fixed to their values at SM-width.

The $gg$- and $qg$-initiated interference contributions comprise simulated events with both positive and negative weights, since the antisymmetric component of the interference changes sign across the Higgs boson mass peak. Because a probability density function must be non-negative everywhere, a dedicated procedure is used to construct a well-behaved analytic model: the interference model in category $c$ is expressed as $f_I^c(m_{\gamma\gamma}) = n_+^c\, f_+^c(m_{\gamma\gamma}) + n_-^c\, f_-^c(m_{\gamma\gamma})$, where $f_+^c$ and $f_-^c$ are asymmetric DSCB functions fitted to the positive-weight and negative-weight subsets of the simulated events, and $n_+^c > 0$ and $n_-^c < 0$ are the MC-predicted yields of each subset, fixed in the fit to data. Although $f_I^c$ itself can take negative values, its components are non-negative probability densities, and the sum of the interference model with the signal and background models ensures that the total probability density entering the likelihood remains non-negative. The interference model is defined separately for the $gg$- and $qg$-initiated contributions, giving two probability density functions $f_{I,gg}^c$ and $f_{I,qg}^c$ used in the final statistical model. The shape parameters of all functions are fixed to their SM-width values, since the dependence of the interference lineshape on $\Gamma_H$ is found to be negligible within the sensitivity of the analysis; the accuracy of this description is assessed with the injection tests described below.

The \mgg distribution of the non-resonant background in each category is described by an analytic function whose parameters are determined from a fit to the data.  The functional form is chosen with the spurious-signal method~\cite{HIGG-2020-16}, which requires the signal yield fitted on a background-only template to be small compared with its statistical uncertainty.  An exponentiated second-order polynomial satisfies this criterion with the fewest degrees of freedom and is selected for both categories.  The background normalisations and shape parameters are free parameters in the final fit.

The main sources of systematic uncertainty are the external constraint on the Higgs boson mass and the photon energy scale on the experimental side, and the uncertainty in the normalisation of the $gg$-interference on the theory side.  They are described below, together with their evaluation and expected impact.

Since the interference modifies the lineshape and in particular shifts the diphoton mass peak, an independent constraint on $m_H$ helps to improve the sensitivity of the analysis.  The Higgs boson mass is constrained to the value measured in the $H\to ZZ^{*}\to 4\ell$ channel using Run 1 and Run 2 data, $m_H = 124.94 \pm 0.18~\GeV$~\cite{HIGG-2020-07}, and the measurement uncertainty is implemented as a single nuisance parameter with a Gaussian constraint in the likelihood function.  On-shell signal--background interference in the $H\to ZZ^{*}$ channel is substantially smaller than in the diphoton channel~\cite{Dixon:2013,Kauer:2012hd}, so any resulting shift in the measured $m_H$ value would be negligible compared with its quoted uncertainty and the external constraint would not be significantly affected.  The corresponding uncertainty is the largest experimental contribution to the total uncertainty in \GH, amounting to 43~\MeV at 68\% confidence level (CL)\@.

The photon energy scale uncertainties affect the measured peak position and thus partially mimic the interference-induced mass shift.  They are modelled with 70 independent components, following the scheme of the Run~2 $H\to\gamma\gamma$ mass measurement~\cite{HIGG-2019-16}, and are propagated to the peak positions of the signal and interference models through the relative variation of the mean of the simulated \mgg distribution under each component.  The energy scale uncertainties are further constrained by the measurement of the linearity of the energy response as a function of the electron transverse energy in $Z\to e^+e^-$ events, performed as part of the energy calibration~\cite{EGAM-2021-02}.  The photon energy resolution uncertainties are described by nine components and are propagated to the model widths through the relative variation of the inter-quartile range of the \mgg distribution.  The largest individual components induce variations of up to $0.07\%$ in the peak positions and up to $5\%$ in the widths of the signal and interference models, with both maxima taken over all components and both \pTgg categories.  In total, the photon energy scale and resolution uncertainties contribute 22~\MeV and 2~\MeV, respectively, to the 68\%~CL upper bound on \GH.

Uncertainties in the photon identification efficiency, the trigger efficiency and the modelling of pile-up affect the expected signal and interference yields by $\pm 1.8\%$, $\pm 0.9\%$ and $\pm 0.6\%$, respectively. The integrated luminosity contributes a global normalisation uncertainty of $0.83\%$ on the signal and interference yields~\cite{DAPR-2021-01}.  Possible biases arising from the choice of the analytic models are quantified with injection tests, in which pseudo-data are constructed at $\GH = 400~\MeV$ with one component in turn (signal, interference or background) taken directly from its simulated template while the remaining components are described by their analytic models, and the fitted value of \GH is compared with the injected one.  The resulting biases are of the order of $5\%$ of the injected value and not statistically significant; they are nonetheless included in the likelihood function as additional nuisance parameters, uncorrelated between categories.

On the theory side, the higher-order QCD correction factor $K_I = 1.64$ applied to the $gg$-interference yield carries a relative uncertainty of $39\%$, covering the difference between the NLO calculation implemented in the simulation and the NNLOsv result~\cite{Bargiela:2023,Buccioni:2026}.  It is implemented as a single nuisance parameter acting on the $gg$-interference normalisation, fully correlated between the two \pTgg categories, and constitutes the largest single contribution to the total uncertainty in \GH, amounting to 49~\MeV at 68\%~CL\@.  No uncertainty is assigned to the signal K-factor $K_S = 1.4$, since its effect on the signal normalisation is fully absorbed by the freely profiled signal-strength parameter $\mu$.  The effect of missing higher-order QCD terms in the signal and interference predictions is evaluated by varying the renormalisation and factorisation scales up and down by a factor of two and taking the envelope of the variations.  For the signal, the resulting yield uncertainties are $\pm 6.3\%$ and $\pm 7.6\%$ in the low- and high-\pTgg categories, respectively.  For the interference processes, whose overall normalisation uncertainty is already covered by the $K_I$ nuisance parameter, only the induced migration of events between the two categories is considered, amounting to $\pm 3.6\%$ ($\pm 3.4\%$) for the $gg$-interference and $\pm 15.8\%$ ($\pm 4.3\%$) for the $qg$-interference in the low-\pTgg (high-\pTgg) category.  Uncertainties from the parton distribution functions, evaluated with the 100 replicas of the NNPDF3.0 set~\cite{Ball:2014uwa} and combined in quadrature with the effect of varying $\alpha_{\mathrm{s}}$, amount to $2.6\%$ for the signal and $gg$-interference yields and up to $3.9\%$ for the $qg$-interference, with a negligible impact on the \mgg shape.  The $H\to\gamma\gamma$ branching fraction carries an uncertainty of $\pm2.9\%$~\cite{deFlorian:2016spz}, implemented as a log-normal constraint on the signal normalisation.  Since \Sherpa provides the only available implementation of the signal and interference processes simultaneously, the samples are generated exclusively with the CSS parton shower. Parton-shower uncertainties can therefore not be assessed by comparing different parton-shower algorithms and are instead evaluated by varying internal settings of the simulation.  The scale factor of the strong coupling used in the initial-state CSS shower is varied up and down by a factor of two relative to its nominal value, which was chosen to optimise the agreement of the signal \pTgg spectrum with the NNLO-plus-parton-shower (NNLOPS) prediction~\cite{Hamilton:2013fea}; the resulting migration of events between the two categories is up to $6\%$ for the signal, $9\%$ for the $gg$-interference and $3\%$ for the $qg$-interference.  The choice of the NLO subtraction scheme, which emulates the difference between the CSS and \AtlasMC{Dire}~\cite{Hoche:2015} parton-shower algorithms, has a negligible effect on the signal and induces migrations of up to $6\%$ for the $gg$-interference and $23\%$ for the $qg$-interference.  These variations affect the relative yields of the two categories and are included as additional nuisance parameters.  The non-interfering contributions from the other Higgs boson production modes are not scaled by $\mu$: they are fixed to their SM expected yields within the corresponding theoretical cross-section and experimental uncertainties.

The Higgs boson total width \GH is determined from a simultaneous maximum-likelihood fit to the \mgg distributions in the $\pTgg < 30~\GeV$ and $\pTgg \geq 30~\GeV$ categories.  The likelihood function depends on \GH and on additional parameters: the signal strength $\mu$, which scales only the $gg$-induced signal component, and the nuisance parameters describing the background models and the systematic uncertainties, including the constraint on the Higgs boson mass.  The test statistic is the profile likelihood ratio~\cite{Rolke:2004mj}.  Upper limits on \GH at 95\%~CL are derived with the CL$_\text{s}$ method~\cite{Junk:1999kv,Read:2002hq}, based on the ratio of one-sided $p$-values $\text{CL}_\text{s} = p_{\GH} / (1 - p_{\text{SM}})$, where $p_{\GH}$ is the $p$-value for the tested value of \GH and $p_{\text{SM}}$ is the $p$-value obtained with the same test statistic under the SM hypothesis ($\GHSM = 4.1~\MeV$, $\mu = 1$, $m_H = 124.94~\GeV$)~\cite{HIGG-2017-06}.  Here $p_{\text{SM}}$ is used as the correction in the denominator of CL$_\text{s}$, rather than the conventional background-only hypothesis, because $\GH = 0$ is not physically meaningful.  The test-statistic distributions used to compute the CL$_\text{s}$ are evaluated with pseudo-experiments. The expected limit is obtained from pseudo-experiments generated under the SM hypothesis.  Figure~\ref{fig:postfit} shows the \mgg distributions of the data in the two categories together with the result of the simultaneous fit.  The observed (expected) upper limit on \GH is $307\ (258)~\MeV$ at 95\%~CL, corresponding to approximately $75\ (63)$ times the SM width.  The best-fit value is $\hat{\Gamma}_H = 13^{+75}_{-13}~\MeV$, consistent with the SM prediction of $4.1~\MeV$. The observed limit exceeds the expected one because the best-fit value is slightly above the SM hypothesis assumed in the expected limit computation.  The signal strength, profiled in the fit, is measured to be $\mu = 1.08 \pm 0.12$, compatible with the SM expectation. When only the statistical uncertainty is considered, the observed (expected) upper limit on \GH is $105\ (87)~\MeV$ at 95\%~CL, corresponding to approximately $26\ (21)$ times the SM width. Figure~\ref{fig:nllscan} shows the observed and expected likelihood scans as a function of \GH, together with the scans obtained when the nuisance parameters associated with the theory uncertainties, and then with all systematic uncertainties, are fixed to their best-fit values, thereby removing the corresponding contributions to the uncertainty in \GH. The confidence levels in Figure~\ref{fig:nllscan} are defined using the asymptotic approximation~\cite{Cowan:2010js}, whose coverage agrees well with that evaluated with pseudo-experiments except for \GH values larger than 250~\MeV, and for values close to $\Gamma_H = 0$, where the physical lower boundary and the square-root dependence of the interference on \GH invalidate the $\chi^2$ approximation for the test-statistic distribution.  The measurement is dominated by systematic uncertainties, with the leading contributions from the theoretical uncertainty in the $gg$-interference normalisation and from the external $m_H$ constraint.  The impact of the $K_I$ normalisation uncertainty grows with \GH, since the interference scales as $\sqrt{\GH}$; at larger values of \GH, close to the 95\%~CL upper limit, it therefore becomes by far the dominant uncertainty.

\begin{figure}[tp]
\centering
\subfloat[]{%
\centering
\includegraphics[width=0.49\linewidth]{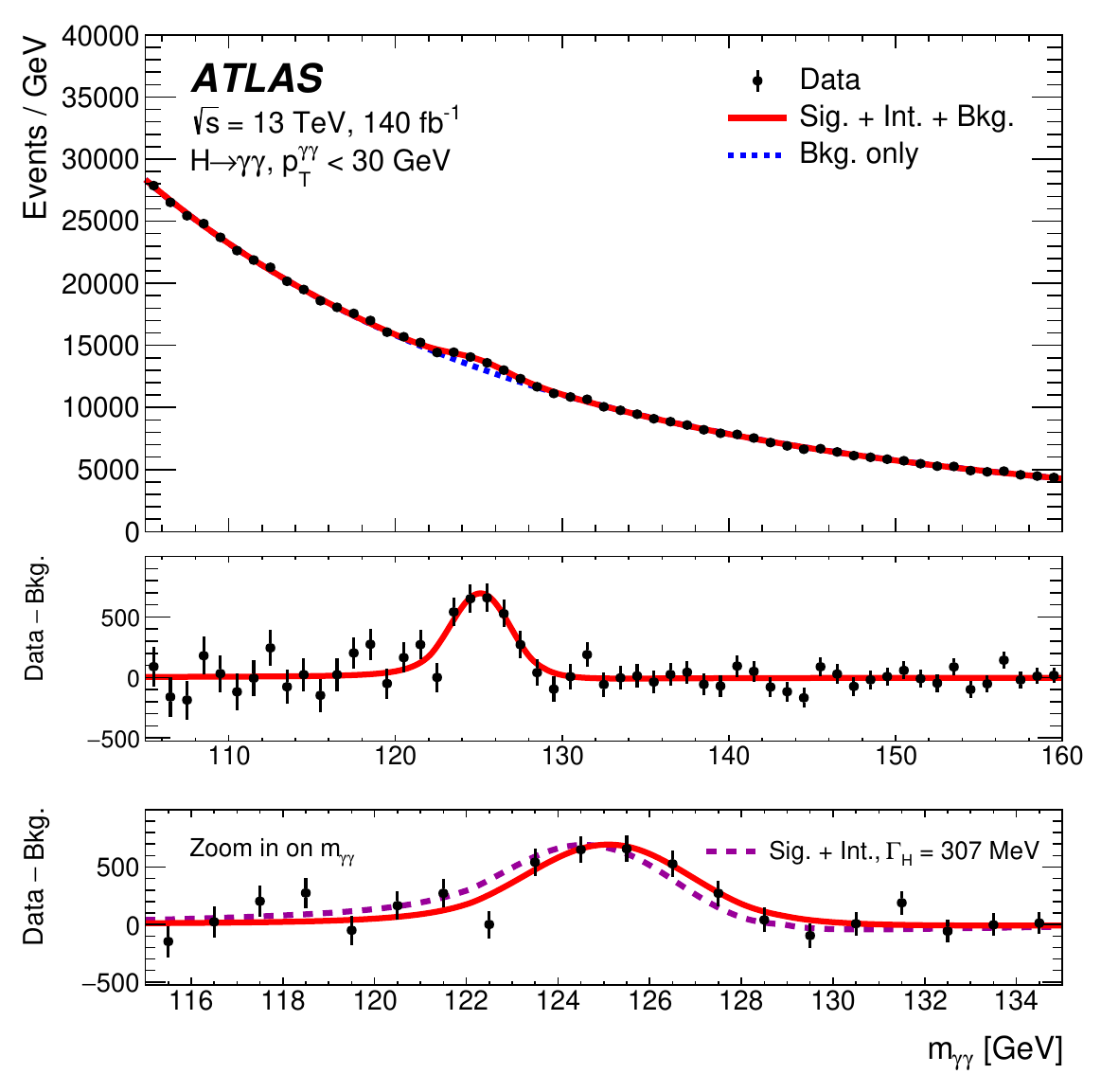}
}
\hfill
\subfloat[]{%
\centering
\includegraphics[width=0.49\linewidth]{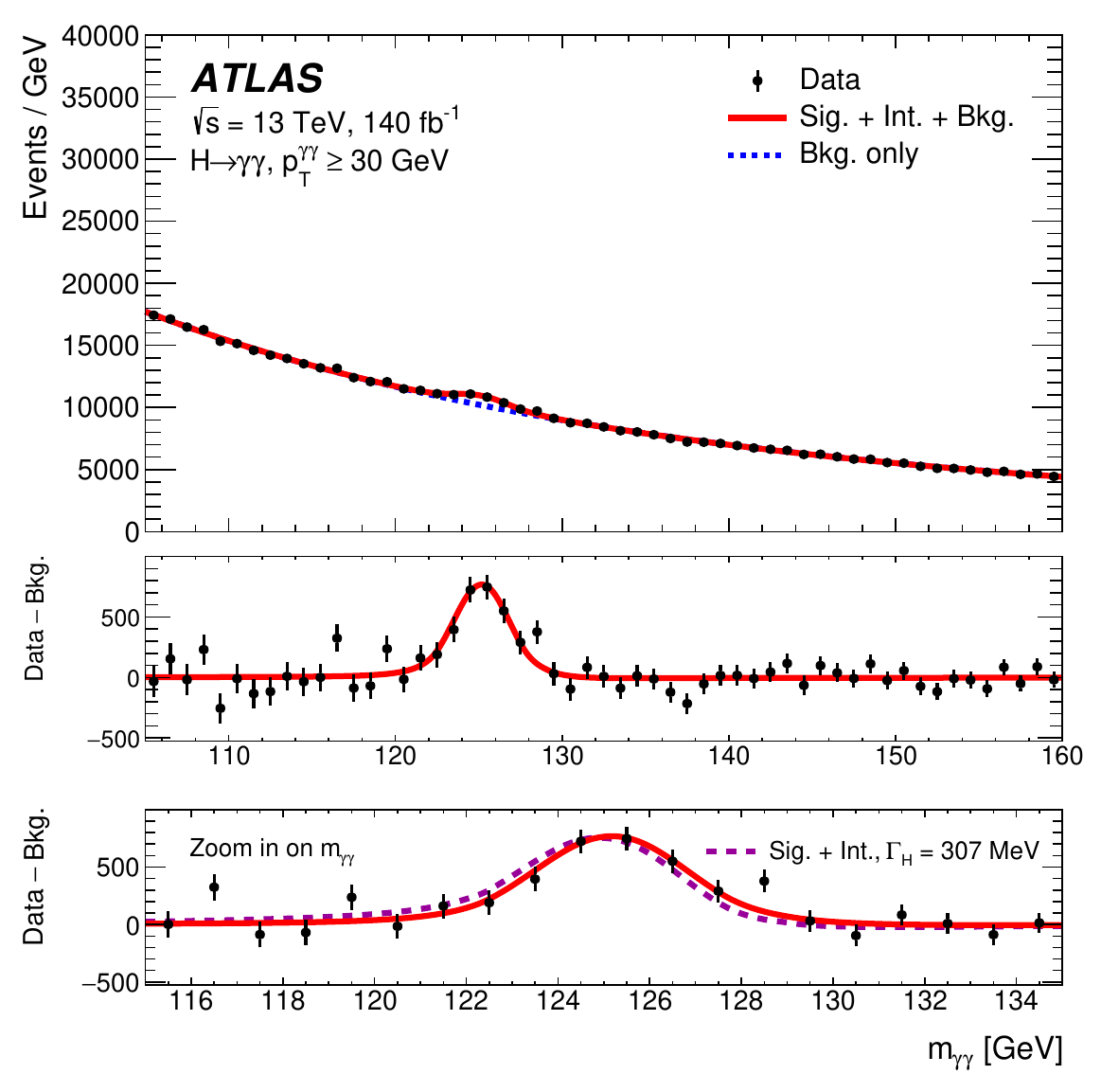}
}
\captionsetup{width=0.9\textwidth}
\caption{Diphoton invariant mass distributions of the selected events in the (a) $\pTgg < 30~\GeV$ and (b) $\pTgg \geq 30~\GeV$ categories.  The data points are shown as black circles with statistical uncertainties.  The solid red curve shows the signal-plus-interference-plus-background model obtained from the simultaneous fit to both categories, evaluated at the best-fit width $\hat{\Gamma}_H = 13~\MeV$; the dotted blue line shows the non-resonant background component of the same fit. The bottom panels show the residuals of the data relative to the non-resonant background component of the fit, together with the corresponding signal-plus-interference components of the two models. In the lowest panel, the dashed purple line shows the signal-plus-interference model evaluated at $\GH = 307~\MeV$, the observed 95\%~CL upper limit.}
\label{fig:postfit}
\end{figure}

\begin{figure}[tp]
\centering
\includegraphics[width=0.6\textwidth]{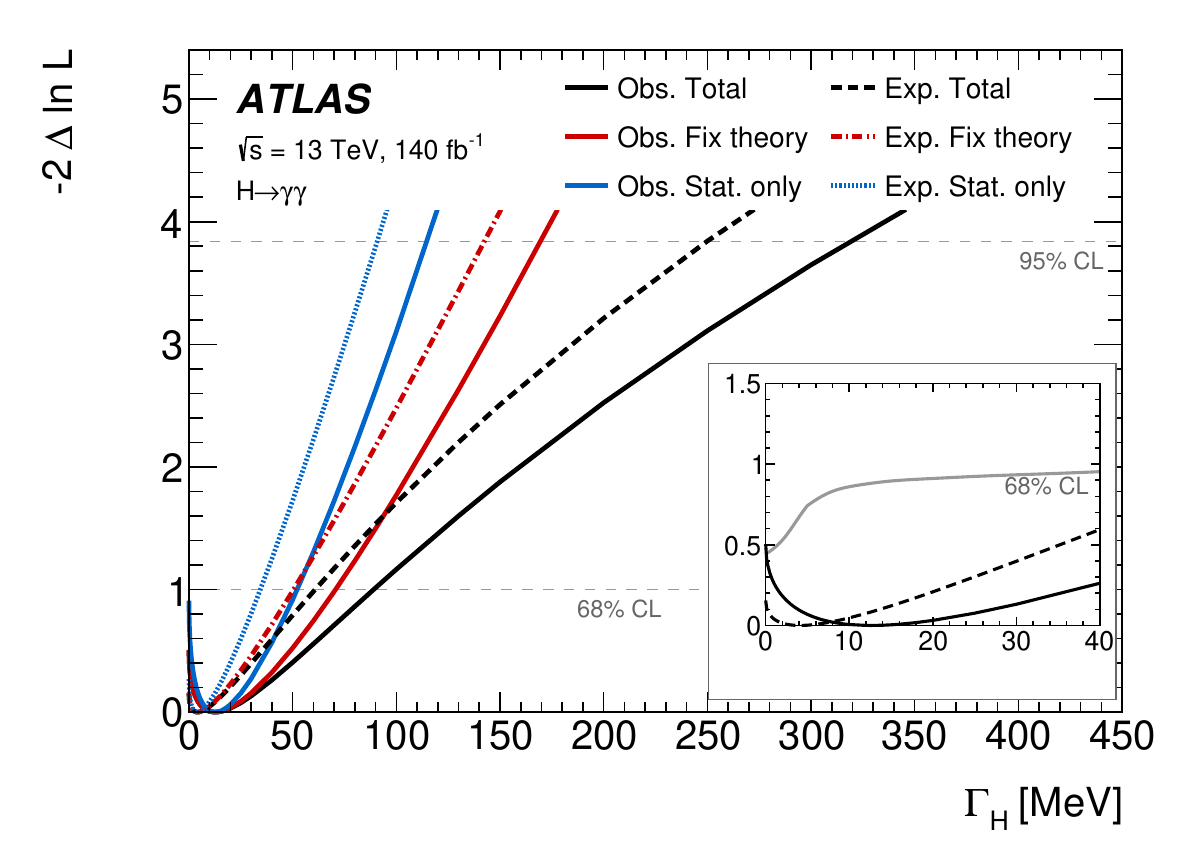}
\captionsetup{width=0.9\textwidth}
\caption{The solid and dashed curves show the observed and expected likelihood scans as a function of $\Gamma_H$. The black lines shows the total likelihood scans.  The additional curves are obtained by fixing the nuisance parameters associated with the theory uncertainties (``fix theory''), and with all systematic uncertainties (``stat.\ only''), to their best-fit values, thereby removing the corresponding uncertainty contributions from the scan. The inset shows a zoomed-in view of the observed and expected total likelihood curves around their minima, for $\Gamma_H \in [0, 40]$~\MeV. The confidence levels in the main canvas are defined using the asymptotic approximation, whose coverage agrees well with that evaluated with pseudo-experiments except for \GH values larger than 250~\MeV, and for values close to $\Gamma_H = 0$; in the inset, the 68\%~CL interval is therefore determined from pseudo-experiments.}
\label{fig:nllscan}
\end{figure}

\FloatBarrier

In summary, the ATLAS experiment reports a direct, on-shell constraint on the total width of the Higgs boson, using the interference between the resonant $gg\to\Hgg$ amplitude and the continuum diphoton background, with a sub-dominant $qg$-initiated interference contribution also included. The measurement uses the full Run~2 $pp$ collision data at $\sqrt{s} = 13~\TeV$, corresponding to an integrated luminosity of $140~\ifb$. The observed 95\%~CL upper limit is $\GH < 307~\MeV$ compared with an expected limit of $\GH < 258~\MeV$. When only statistical uncertainties are considered, the observed (expected) limit improves to $105\ (87)~\MeV$.  The \GH constraint at the 95\%~CL is limited mainly by the theoretical uncertainty in the normalisation of the $gg$-interference, motivating improved calculations of the interference, including matching to parton showers, for future analyses. While the sensitivity is about 30 times weaker than that of the indirect off-shell determinations~\cite{HIGG-2018-32,HIGP-2024-14,HIGP-2024-05,CMS-HIG-21-013,CMS-HIG-21-019,CMS-HIG-24-011}, the present measurement does not rely on theoretical assumptions such as the equality of the on-shell and off-shell Higgs boson couplings: the sensitivity to \GH arises only from the interference-induced distortion of the \mgg lineshape, together with the residual effect of the interference on the yields in the two categories.  The two approaches therefore provide complementary probes of the Higgs boson total width and have different sensitivities to new physics.

\section*{Acknowledgements}

%
%

%
%

We thank CERN for the very successful operation of the LHC and its injectors, as well as the support staff at
CERN and at our institutions worldwide without whom ATLAS could not be operated efficiently.

The crucial computing support from all WLCG partners is acknowledged gratefully, in particular from CERN, the ATLAS Tier-1 facilities at TRIUMF/SFU (Canada), NDGF (Denmark, Norway, Sweden), CC-IN2P3 (France), KIT/GridKA (Germany), INFN-CNAF (Italy), NL-T1 (Netherlands), PIC (Spain), RAL (UK) and BNL (USA), the Tier-2 facilities worldwide and large non-WLCG resource providers. Major contributors of computing resources are listed in Ref.~\cite{ATL-SOFT-PUB-2026-001}.

We gratefully acknowledge the support of ANPCyT, Argentina; YerPhI, Armenia; ARC, Australia; BMWFW and FWF, Austria; ANAS, Azerbaijan; CNPq and FAPESP, Brazil; NSERC, NRC and CFI, Canada; CERN; ANID, Chile; CAS, MOST, NSFC and FRFCU, China; Minciencias, Colombia; MEYS CR, Czech Republic; DNRF and DNSRC, Denmark; IN2P3-CNRS and CEA-DRF/IRFU, France; SRNSFG, Georgia; BMFTR, HGF and MPG, Germany; GSRI, Greece; RGC and Hong Kong SAR, China; ICHEP and Academy of Sciences and Humanities, Israel; INFN, Italy; MEXT and JSPS, Japan; CNRST, Morocco; NWO, Netherlands; RCN, Norway; MNiSW, Poland; FCT, Portugal; MNE/IFA, Romania; MSTDI, Serbia; MSSR, Slovakia; ARIS and MVZI, Slovenia; DSI/NRF, South Africa; MICIU/AEI, Spain; SRC and Wallenberg Foundation, Sweden; SERI, SNSF and Cantons of Bern and Geneva, Switzerland; NSTC, Taipei; TENMAK, T\"urkiye; STFC/UKRI, United Kingdom; DOE and NSF, United States of America.

Individual groups and members have received support from BCKDF, CANARIE, CRC and DRAC, Canada; CERN-CZ and PRIMUS, Czech Republic; COST, ERC, ERDF, Horizon 2020 and Marie Sk{\l}odowska-Curie Actions, European Union; Investissements d'Avenir Labex, Investissements d'Avenir Idex and ANR, France; DFG and AvH Foundation, Germany; Herakleitos, Thales and Aristeia programmes co-financed by EU-ESF and the Greek NSRF, Greece; BSF-NSF and MINERVA, Israel; NCN and NAWA, Poland; La Caixa Banking Foundation, CERCA and AGAUR programs from Generalitat de Catalunya and PROMETEO and GenT Programmes Generalitat Valenciana, Spain; G\"{o}ran Gustafssons Stiftelse, Sweden; The Royal Society and Leverhulme Trust, United Kingdom; Eric and Wendy Schmidt Fund for Strategic Innovation, United States of America.

In addition, individual members wish to acknowledge support from Chile: Agencia Nacional de Investigaci\'on y Desarrollo (ANID FONDECYT reg. 1230987, FONDECYT 1230812, FONDECYT 1240864, Fondecyt 3240661, Fondecyt Regular 1240721); China: Fundamental Research Funds for the Central Universities (010-63263105), Chinese Ministry of Science and Technology (MOST-2023YFA1605700, MOST-2023YFA1609300), National Natural Science Foundation of China (NSFC 12275265, NSFC-W2543005); Czech Republic: Czech Science Foundation (GACR - 24-11373S), Ministry of Education Youth and Sports (ERC-CZ-LL2327, FORTE CZ.02.01.01/00/22\_008/0004632); EU: H2020 European Research Council (ERC - 101002463); European Union: European Research Council (BARD No. 101116429, ERC - 101219398, ERC 101089007), European Regional Development Fund (HE COFUND GA No.101081355, ERDF), Marie Sklodowska-Curie Actions (GAP-101168829); France: Agence Nationale de la Recherche (ANR-22-EDIR-0002, ANR-24-CE31-0504-01); China: Research Grants Council (GRF); Italy: Istituto Nazionale di Fisica Nucleare (LHC-MIUR - 28003/2025), Ministero dell'Università e della Ricerca (NextGenEU I53D23000820006 M4C2.1.1, SOE2024\_0000023); Japan: Japan Society for the Promotion of Science (JSPS KAKENHI  JP25H0063, JSPS KAKENHI JP22H04944, JSPS KAKENHI JP24K23939, JSPS KAKENHI JP24KK0251, JSPS KAKENHI JP25H00650, JSPS KAKENHI JP25H01291, JSPS KAKENHI JP25K01011, JSPS KAKENHI JP25K01023, JSPS KAKENHI JP25KK0047); Poland: Polish National Science Centre (NCN 2021/42/E/ST2/00350, NCN OPUS 2023/51/B/ST2/02507, NCN OPUS nr 2022/47/B/ST2/03059, NCN UMO-2019/34/E/ST2/00393, UMO-2022/47/O/ST2/00148, UMO-2023/49/B/ST2/04085, UMO-2023/51/B/ST2/00920, UMO-2024/53/N/ST2/00869); Spain: Agència de Gestió d'Ajuts Universitaris i de Recerca. (AGAUR - 2023 BP 00141), Ministry of Science and Innovation (RYC2019-028510-I, RYC2020-030254-I, RYC2021-031273-I, RYC2022-038164-I), Ministerio de Ciencia, Innovación y Universidades/Agencia Estatal de Investigaci\'on (PID2022-142604OB-C22); Sweden: Carl Trygger Foundation (Carl Trygger Foundation CTS 22:2312), Swedish Research Council (Swedish Research Council 2023-04654, VR 2021-03651, VR 2022-03845, VR 2022-04683, VR 2023-03403, VR 2024-05451, VR 2025-05940), Knut and Alice Wallenberg Foundation (KAW 2023.0366); United Kingdom: The Binks Trust, Royal Society (NIF-R1-231091); United States of America: U.S. Department of Energy (ECA DE-AC02-76SF00515), John Templeton Foundation (John Templeton Foundation 63206), Neubauer Family Foundation.

%
%


\printbibliography

\clearpage
 
\begin{flushleft}
\hypersetup{urlcolor=black}
{\Large The ATLAS Collaboration}

\bigskip

\AtlasOrcid[0000-0002-6665-4934]{G.~Aad}$^\textrm{\scriptsize 102}$,
\AtlasOrcid[0000-0001-7616-1554]{E.~Aakvaag}$^\textrm{\scriptsize 17}$,
\AtlasOrcid[0000-0002-5888-2734]{B.~Abbott}$^\textrm{\scriptsize 121}$,
\AtlasOrcid[0000-0002-0287-5869]{S.~Abdelhameed}$^\textrm{\scriptsize 83b}$,
\AtlasOrcid[0000-0002-1002-1652]{K.~Abeling}$^\textrm{\scriptsize 54}$,
\AtlasOrcid[0000-0001-5763-2760]{N.J.~Abicht}$^\textrm{\scriptsize 48}$,
\AtlasOrcid[0000-0002-8496-9294]{S.H.~Abidi}$^\textrm{\scriptsize 30}$,
\AtlasOrcid[0009-0003-6578-220X]{M.~Aboelela}$^\textrm{\scriptsize 44}$,
\AtlasOrcid[0000-0002-9987-2292]{A.~Aboulhorma}$^\textrm{\scriptsize 36e}$,
\AtlasOrcid[0000-0001-5329-6640]{H.~Abramowicz}$^\textrm{\scriptsize 154}$
% ...
\bigskip
\\

% ...
$^{17}$Department for Physics and Technology, University of Bergen, Bergen; Norway.\\
% ...
$^{30}$Physics Department, Brookhaven National Laboratory, Upton NY; United States of America.\\
% ...
$^{36}$$^{(a)}$Facult\'e des Sciences Ain Chock, Universit\'e Hassan II de Casablanca;$^{(b)}$Facult\'{e} des Sciences, Universit\'{e} Ibn-Tofail, K\'{e}nitra;$^{(c)}$Facult\'e des Sciences Semlalia, Universit\'e Cadi Ayyad, LPHEA-Marrakech;$^{(d)}$LPMR, Facult\'e des Sciences, Universit\'e Mohamed Premier, Oujda;$^{(e)}$Facult\'e des sciences, Universit\'e Mohammed V, Rabat;$^{(f)}$Institute of Applied Physics, Mohammed VI Polytechnic University, Ben Guerir; Morocco.\\
% ...
$^{44}$Physics Department, Southern Methodist University, Dallas TX; United States of America.\\
% ...
$^{48}$Fakult\"{a}t Physik, Technische Universit{\"a}t Dortmund, Dortmund; Germany.\\
% ...
$^{54}$II. Physikalisches Institut, Georg-August-Universit\"{a}t G\"ottingen, G\"ottingen; Germany.\\
% ...
$^{83}$$^{(a)}$Khalifa University of Science and Technology, Abu Dhabi;$^{(b)}$New York University Abu Dhabi, Abu Dhabi;$^{(c)}$United Arab Emirates University, Al Ain;$^{(d)}$University of Sharjah, Sharjah; United Arab Emirates.\\
% ...
$^{102}$CPPM, Aix-Marseille Universit\'e, CNRS/IN2P3, Marseille; France.\\
% ...
$^{121}$Homer L. Dodge Department of Physics and Astronomy, University of Oklahoma, Norman OK; United States of America.\\
% ...
$^{154}$Raymond and Beverly Sackler School of Physics and Astronomy, Tel Aviv University, Tel Aviv; Israel.\\
% ...
$^{a}$ Also at Affiliated with an institute formerly covered by a cooperation agreement with CERN.\\
$^{b}$ Also at An-Najah National University, Nablus; Palestine.\\
$^{c}$ Also at Borough of Manhattan Community College, City University of New York, New York NY; United States of America.\\
$^{d}$ Also at Center for Interdisciplinary Research and Innovation (CIRI-AUTH), Thessaloniki; Greece.\\
$^{e}$ Also at Centre of Physics of the Universities of Minho and Porto (CF-UM-UP); Portugal.\\
$^{f}$ Also at CERN, Geneva; Switzerland.\\
% ...

\end{flushleft}


\end{document}